\documentclass{aamod}

\usepackage{natbib}
\bibpunct{(}{)}{;}{a}{}{,}
\usepackage{graphicx}
\usepackage{txfonts}

\begin{document}

\title{Milne-Eddington inversions of the \ion{He}{i} 10830 {\AA}
Stokes profiles: Influence of the Paschen-Back effect}

\author{C. Sasso \and A. Lagg \and S. K. Solanki}

\offprints{C. Sasso, \email{sasso@mps.mpg.de}}

\institute{Max-Planck-Institut f\"ur Sonnensystemforschung, Max-Planck-Str. 2,
Katlenburg-Lindau, Germany}

\date{}

\abstract{The Paschen-Back effect influences the Zeeman sublevels of the
\ion{He}{i} multiplet at 10830 {\AA}, leading to changes in strength and in
position of the Zeeman components of these lines.}{We illustrate the
relevance of this effect using synthetic Stokes profiles of the \ion{He}{i}
10830 {\AA} multiplet lines and investigate its influence on the inversion of
polarimetric data.}{We invert data obtained with the Tenerife Infrared
Polarimeter (TIP) at the German Vacuum Tower Telescope (VTT). We compare the
results of inversions based on synthetic profiles calculated with and without
the Paschen-Back effect being included.}{We find that when taking into
account the incomplete Paschen-Back effect, on average 16\% higher field
strength values are obtained. We also show that this effect is not the main
cause for the area asymmetry exhibited by many \ion{He}{i} 10830 Stokes
$V$-profiles. This points to the importance of velocity and magnetic field
gradients over the formation height range of these lines.}{}


\titlerunning{Inversions of the \ion{He}{i} 10830 {\AA}
Stokes profiles: The Paschen-Back effect}

\maketitle

\section{Introduction}

\begin{figure*}
\centering
\includegraphics[width=17cm]{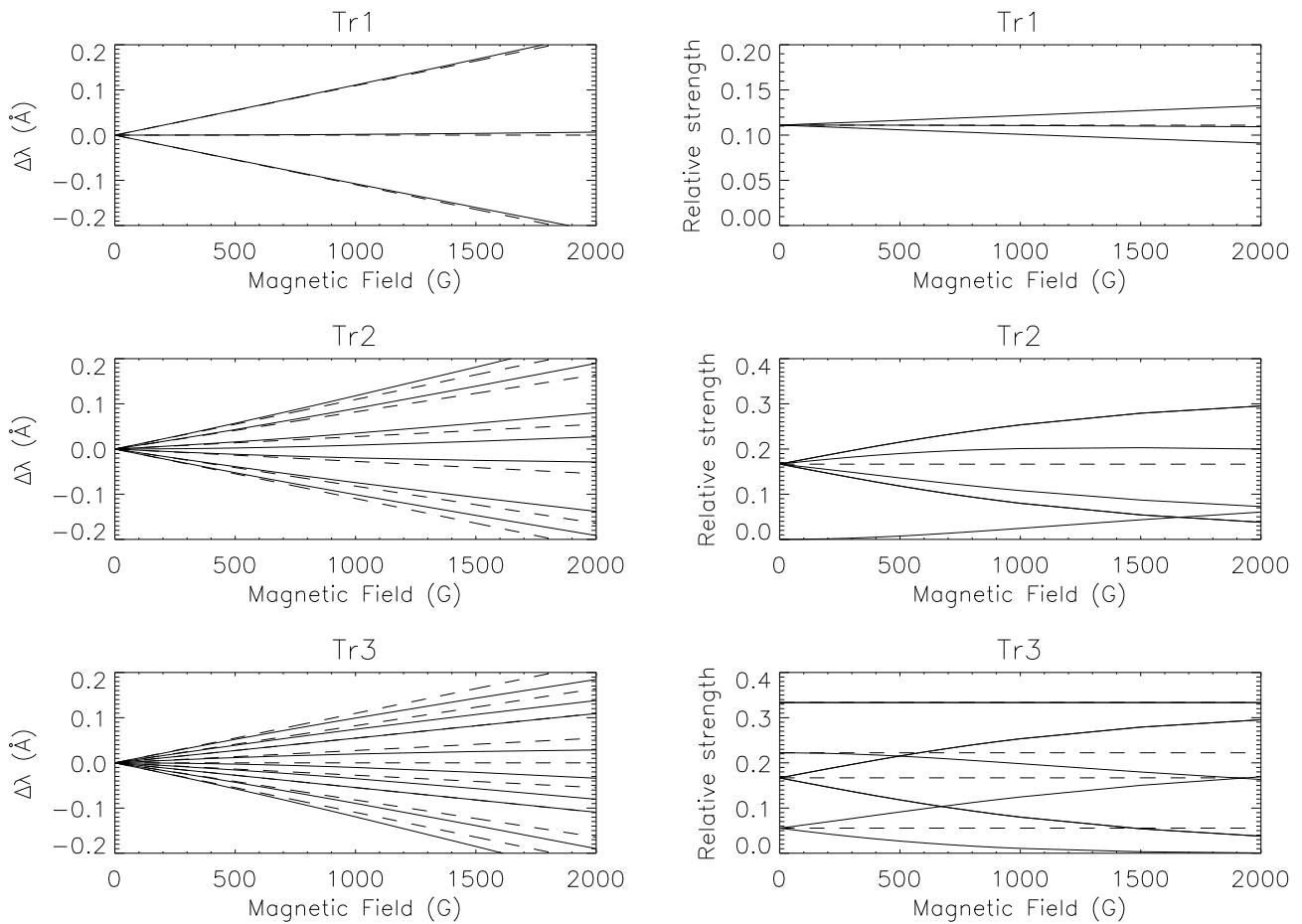}
\caption{Positions and strengths of the Zeeman components as a function of
the magnetic field. The solid lines represent calculations considering the
incomplete Paschen-Back splitting (IPBS), while the dashed lines represent
those using the linear Zeeman splitting (LZS) approximation. In the IPBS
case, the Zeeman components exhibit asymmetric displacement and strengths.
This figure has been produced using data kindly provided by Socas-Navarro
et al. \citep[see Fig.~1 of][]{socasb}.}
\label{fig:split}
\end{figure*}

Spectropolarimetry in the \ion{He}{i} triplet at 10830 {\AA} has become one
of the key tools to determine the magnetic field vector in the upper
chromosphere \citep{Trujillo_nature,solanki_nature}. A reliable determination
of the magnetic field in this region is important for obtaining a better
understanding of the coronal heating mechanism and the coupling between the
relatively cool photosphere and the hot corona. \\
The \ion{He}{i} 10830 {\AA} multiplet originates between the atomic levels
$2^3S_1$ and $2^3P_{2,1,0}$. It comprises a `blue' component at 10829.09
{\AA} with $J_u=0$ (Tr1), and two `red' components at 10830.25 {\AA} with
$J_u=1$ (Tr2) and at 10830.34 {\AA} with $J_u=2$ (Tr3) which are blended
at solar chromospheric temperatures. \\
In previous papers \citep{solanki_nature,lagg} the analysis of the observed
polarization in the \ion{He}{i} 10830 {\AA} multiplet for the determination
of the magnetic field vector was carried out considering the linear Zeeman
splitting (LZS) approximation. \citet{socasa} demonstrated that the magnetic
field vector from spectropolarimetry in the \ion{He}{i} 10830 {\AA} multiplet
must be determined by considering the wavelength positions and the strengths
of the Zeeman components in the incomplete Paschen-Back effect regime. They
demonstrated that neglecting the Paschen-Back effect results in
significant errors in the calculation of its polarization profiles.
For example, \citet{socasa} pointed out that the Paschen-Back effect
produces a $\sim25\%$ reduction in the amplitude of the red component, which
implies that the linear Zeeman effect theory underestimates the magnetic field
strength. For a detailed presentation of the theory of the Zeeman and the
Paschen-Back effects see, e.g., \citet{landi}. \\
Fig.~\ref{fig:split} shows positions and strengths of the Zeeman components
as a function of the magnetic field. The calculations are performed in two
different ways, by using the LZS approximation or taking into account the
incomplete Paschen-Back splitting (IPBS). The figure, produced using data
kindly provided by Socas-Navarro et al., clearly shows that positions of the
transitions are strongly influenced by the Paschen-Back effect, especially
for magnetic field strengths above a few hundred Gauss. The relative
strengths of the transitions also depend on the field strength. The shift in
wavelength position and the asymmetric change of the strengths between the
blue and the red Zeeman sublevels introduce asymmetries in the resulting
Stokes $Q$, $U$ and $V$ profiles. In their paper \citet{socasa} show that
this asymmetry is strong enough to produce a measurable effect in the
observations. \\
Here we present systematic tests of the influence of the Paschen-Back effect
on parameters retrieved from Stokes profile observations of the \ion{He}{i}
triplet. We pay particular attention to the asymmetry of the Stokes
$V$-profile. This parameter is often used to diagnose combined gradients of
the velocity and the magnetic field along the line of sight (LOS), but the
IPBS also introduces an asymmetry into the Stokes $V$-profiles. It is
therefore of considerable interest to see if the asymmetry seen in many
Stokes $V$-profiles of the \ion{He}{i} 10830 {\AA} triplet is due to the
IPBS, or if this line parameter provides information on the structure of the
upper solar chromosphere.

\section{Methods and results}

\begin{figure*}
\centering
\includegraphics[width=8.5cm]{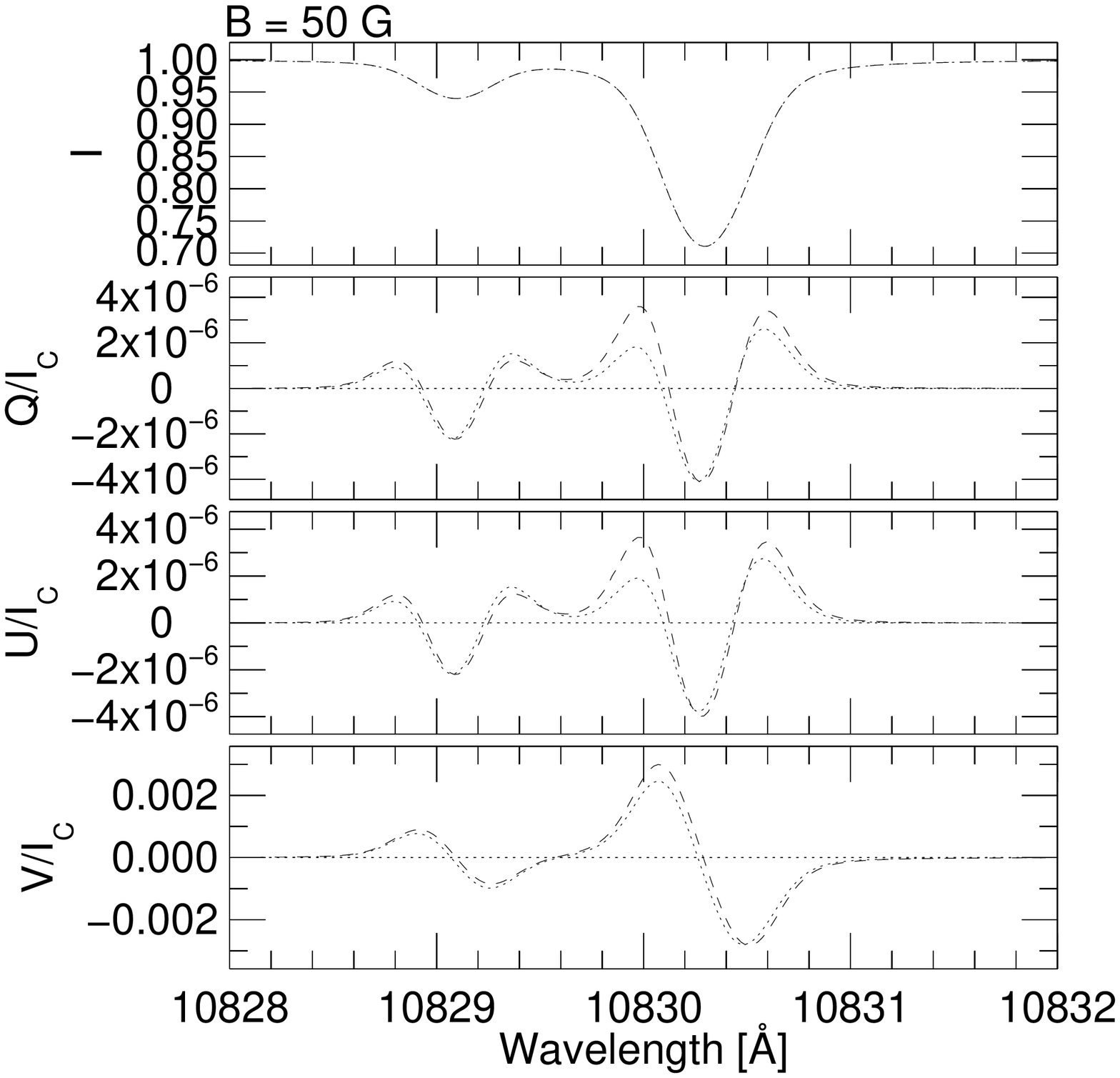}
\includegraphics[width=8.5cm]{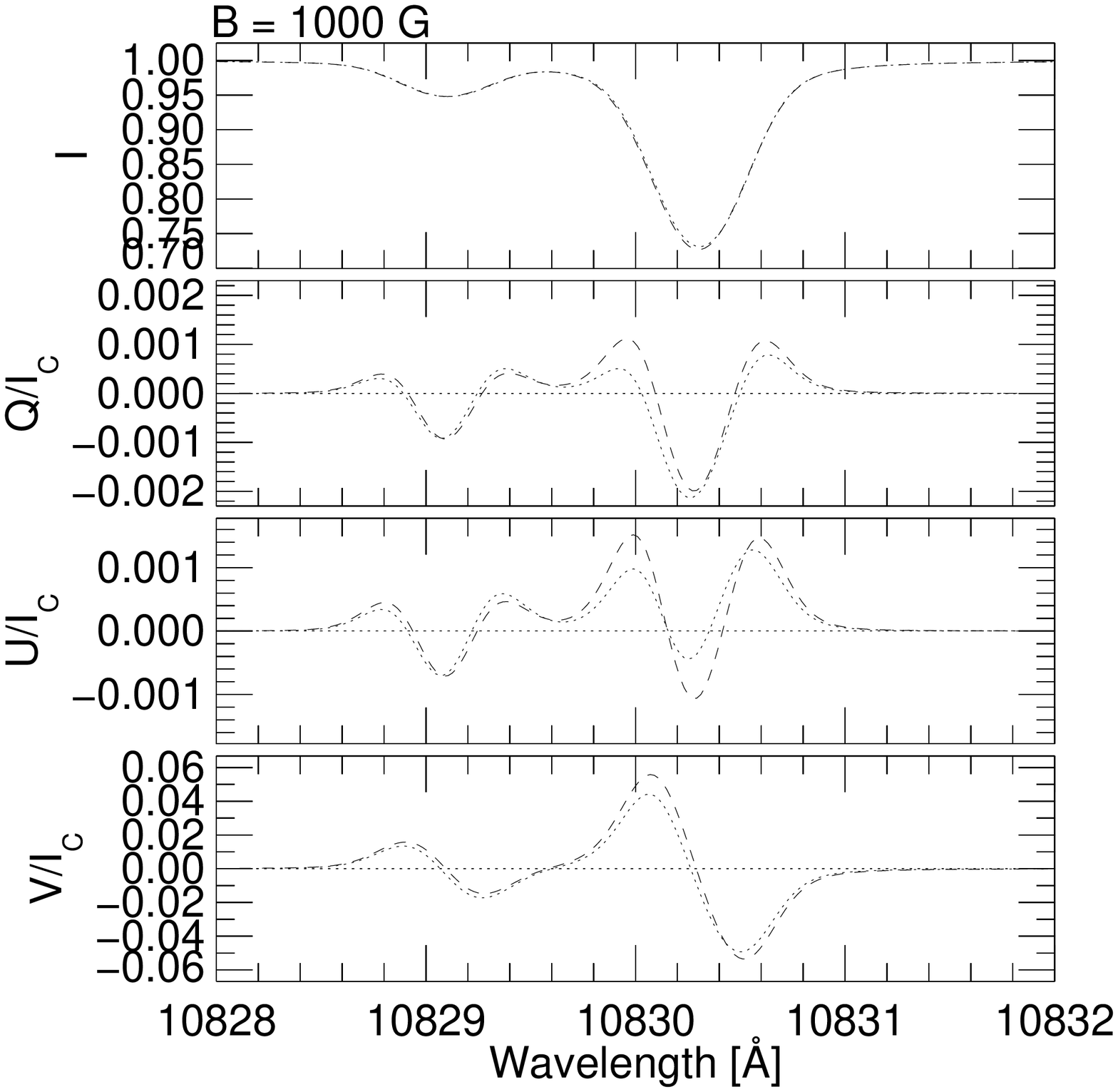}
\caption{Synthetic Stokes profiles of the \ion{He}{i} 10830 {\AA} triplet,
assuming a 50 G (left) and a 1000 G (right) field, respectively, inclined by
30\degr with respect to the line of sight. The Milne-Eddington
parameters we have used for this synthesis are: $\chi=22.5$\degr,
$v_{\textmd{\tiny{LOS}}}=0.0$ m/s, $a=0.3$,
$\Delta\lambda_{\textmd{\tiny{D}}}=0.2$ {\AA}, $\eta_0=5.0$ and $\mu S_1=1.0$.
\citep[For the notation, see][]{lagg}. The dotted profile is obtained by considering
the IPBS, while the dashed one is obtained in the LZS approximation.
The reference direction for Stokes $Q$ is along the $+y$-axis of the
maps in Fig.~\ref{fig:obs_int}.}
\label{fig:pb_nopb}
\end{figure*}

In order to calculate Zeeman components and strengths of the \ion{He}{i}
10830 {\AA} multiplet in the IPBS regime, we have improved the numerical code
for the synthesis and inversion of Stokes profiles in a Milne-Eddington
atmosphere \citep{lagg}, by using the polynomial approximants as proposed by
\citet{socasb}. \\
Fig.~\ref{fig:pb_nopb} compares the synthetic Stokes profiles of the
\ion{He}{i} 10830 {\AA} multiplet for a 50 G and a 1000 G field, inclined by
$30\degr$ with respect to the line of sight, obtained considering LZS only
and including the IPBS. The difference between the profiles obtained when
considering the IPBS (dotted line) or the LZS (dashed line) approximation, is
clearly evident. For example, the $Q$ and $U$ profiles are symmetric and the
$V$ profiles antisymmetric in the LZS case, while they are all asymmetric in
the presence of IPBS. Moreover, if we compare the two panels we can see that
the absolute difference between the profiles calculated using IPBS and LZS,
increases with increasing magnetic field strength although the relative
difference remains roughly the same (note the different vertical scales in
the left and right parts of the figure). Note that in these tests we are
using an effective Land\'e factor for each line of the multiplet for the LZS
case, instead of the calculation of all Zeeman components separately, because
this returns nearly identical results and a considerable gain in computation
time \citep{lagg}.

\subsection{Inversion of synthetic profiles}

\begin{figure*}
\centering
\includegraphics[width=8.5cm,clip=true]{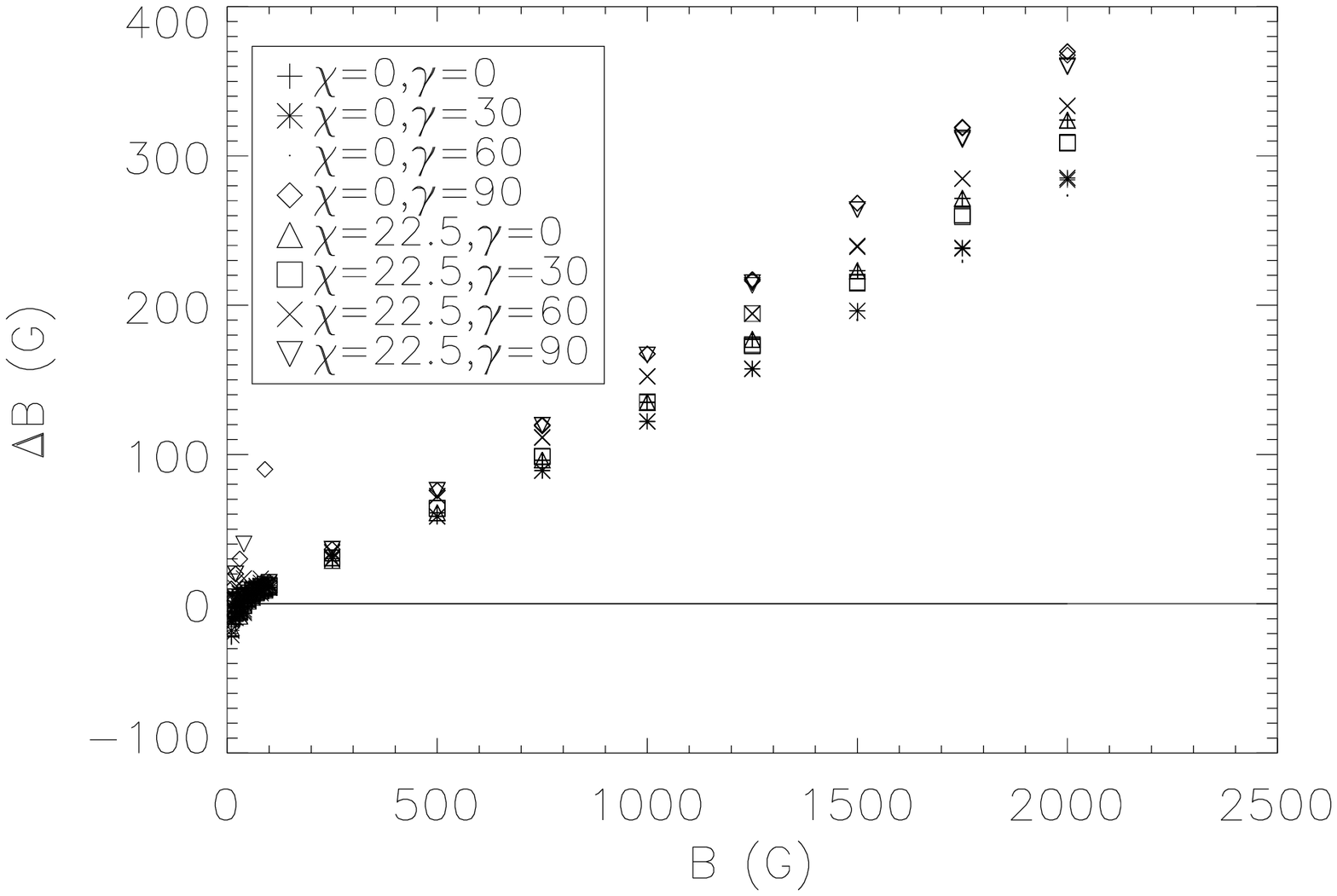}
\includegraphics[width=8.5cm,clip=true]{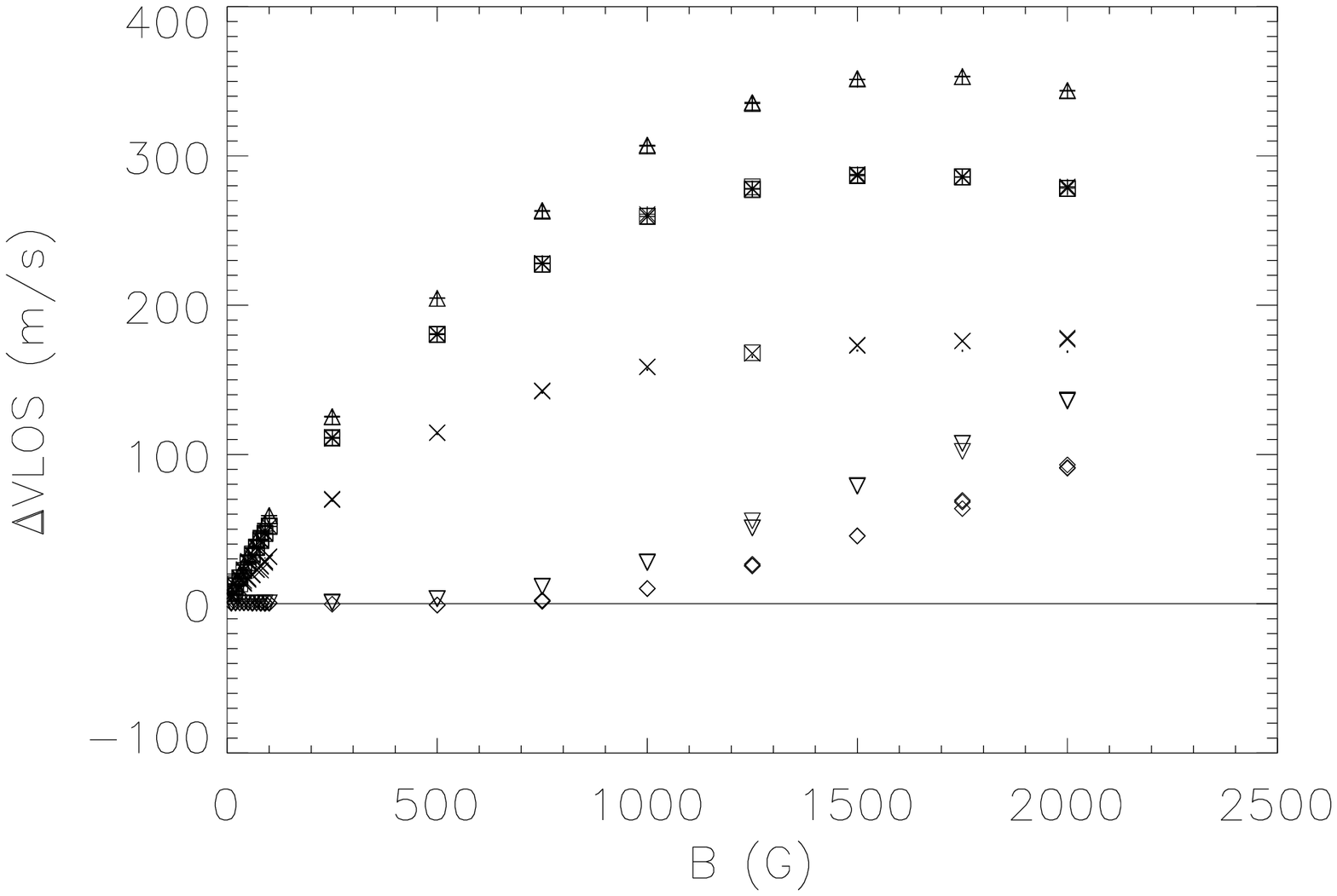}
\caption{On the left: Difference between the values of the magnetic field
strength for a synthetic profile computed including IPBS and the ones
obtained from the LZS inversion of this synthetic IPBS profile ($\Delta$B),
as a function of the magnetic field strength of the synthetic profile for
different inclination ($\gamma$) and azimuthal angles ($\chi$) of the
magnetic field vector (in degrees). On the right: The same for the
LOS-velocity. The retrieved values for magnetic field (left) and LOS-velocity
(right) deviate significantly from the correct values (horizontal solid
line).}
\label{fig:pa}
\end{figure*}

In a next step we analysed the effect of the IPBS on synthetic Stokes
profiles by estimating the error we make when retrieving the values of the
physical parameters (such as the magnetic field vector, the LOS-velocity,
etc.) by using LZS instead of IPBS. For this purpose we calculated synthetic
profiles including IPBS, which are taken to represent an observation. We then
inverted these IPBS profiles assuming LZS only. The difference between the
original magnetic field and velocity values and the values retrieved from the
LZS inversion are displayed in Fig.~\ref{fig:pa}. We see, in both panels, a
significant deviation of the retrieved values from the correct ones,
indicated by the solid line at ordinate 0. More precisely, the error
introduced by neglecting the IPBS increases with the strength of the magnetic
field for both parameters and, for the LOS-velocity, it decreases with
increasing inclination $\gamma$ of the magnetic field to the LOS. Neglecting
IPBS gives artificial downflows of up to 350 m/s. On average the field
strength is underestimated by 16\% if IPBS is neglected, with the exact
fraction lying in the range 14-18\% depending on the magnetic field's
inclination $\gamma$ and its azimuth, $\chi$. \\
The same analysis for the inclination and the azimuthal angle of the magnetic
field shows that these two parameters are accurately retrieved by inversions
even in the LZS approximation.

\subsection{Inversion of observational data}

\begin{figure*}
\centering
\includegraphics[height=8.5cm,angle=90,clip=true]{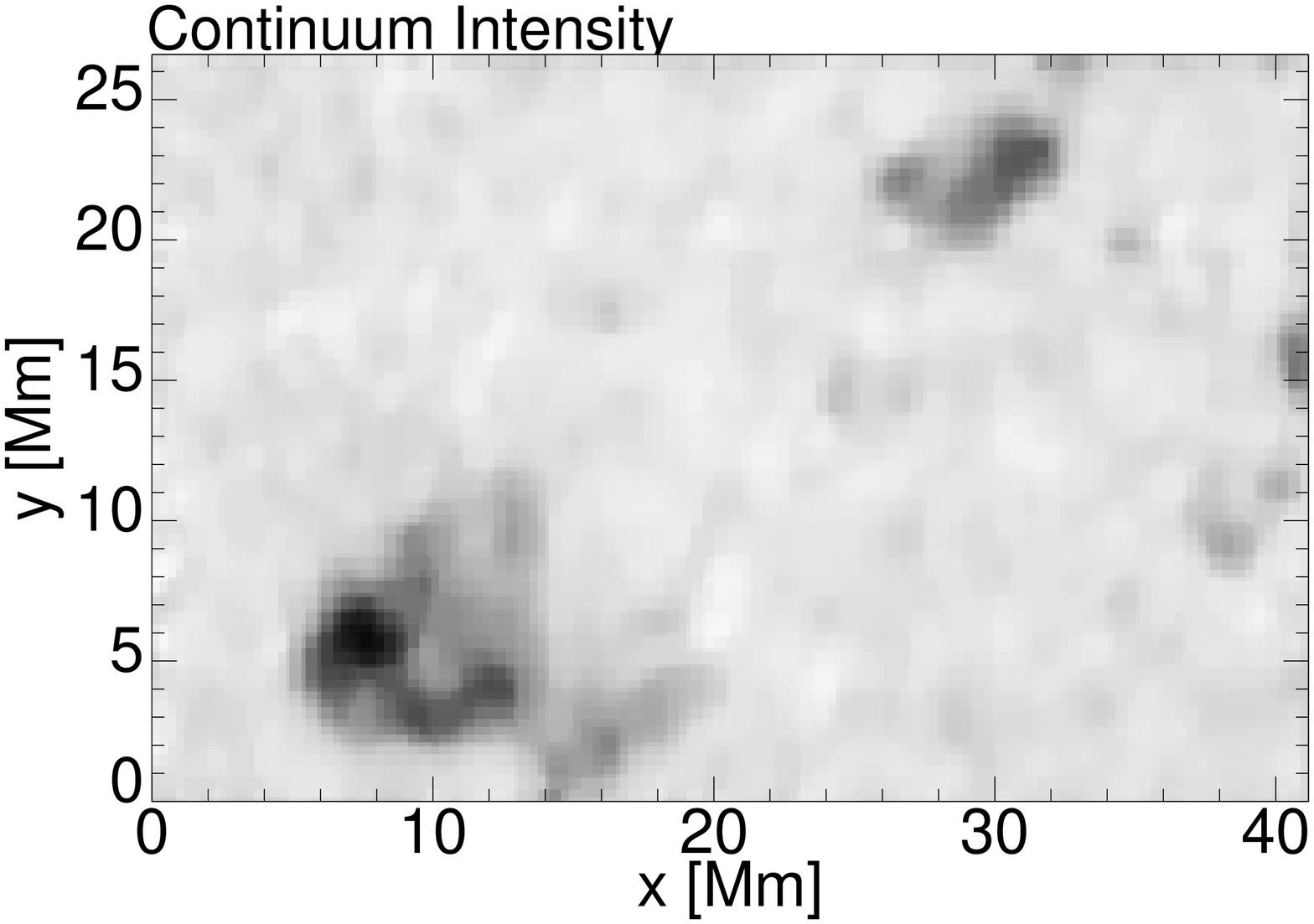}
\includegraphics[height=8.5cm,angle=90,clip=true]{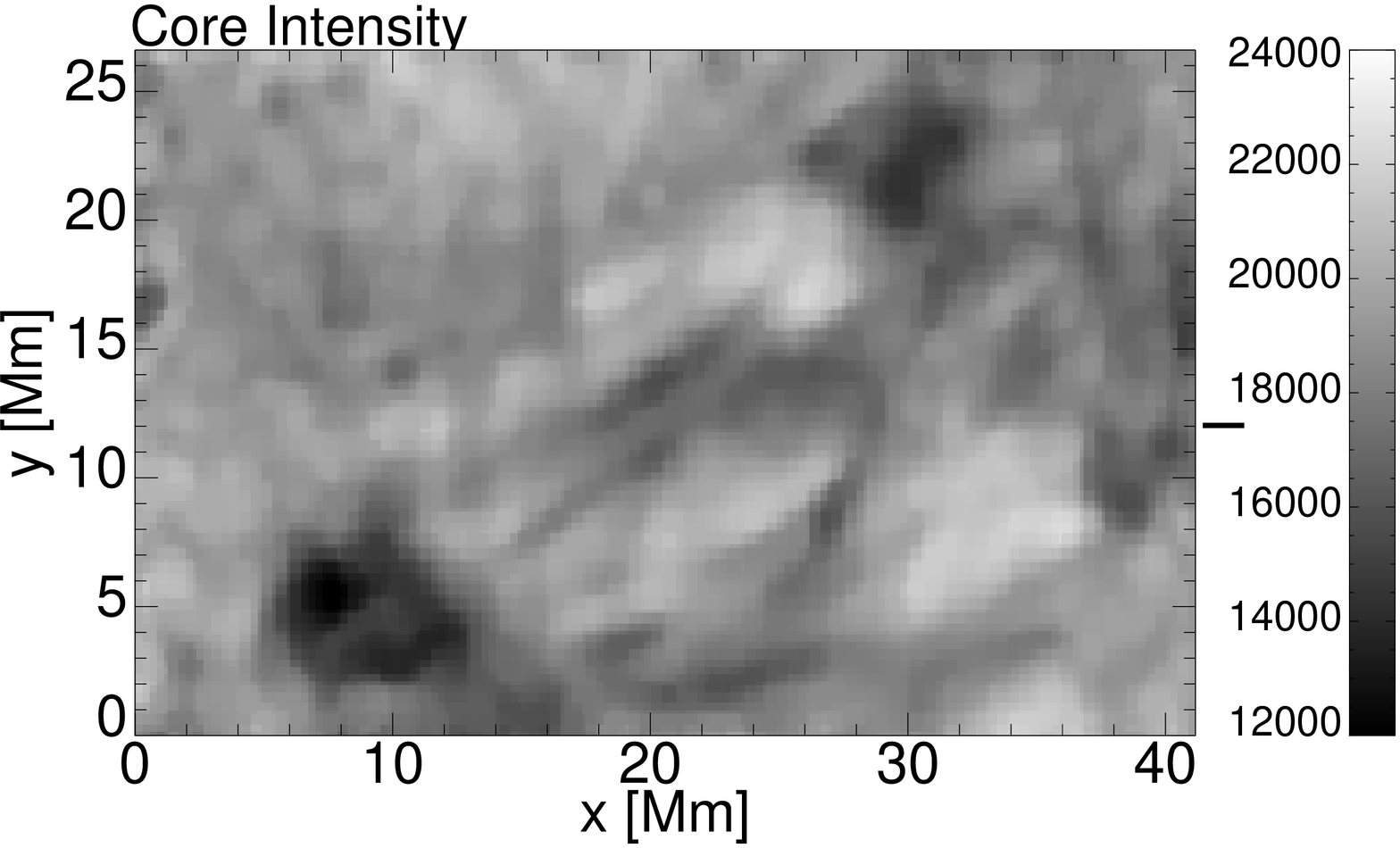}
\caption{Continuum (left) and He core (right) intensity maps of an emerging
flux region (NOAA active region 9451, 33\degr W, 22\degr S)}
\label{fig:obs_int}
\end{figure*}
\begin{figure}
\resizebox{\hsize}{!}{\includegraphics{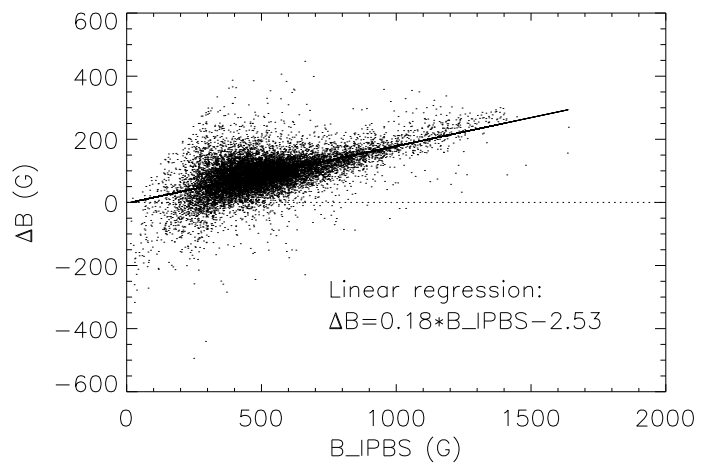}}
\caption{Influence of IPBS on the retrieval of the magnetic field strength
for NOAA 9451. Difference in field strength, $\Delta$B, deduced from
inversions including the IPBS, B$_{\mathrm{IPBS}}$, and those without it,
versus B$_{\mathrm{IPBS}}$. The solid line represents a linear regression.
The larger scatter present in the figure for B$_{\mathrm{IPBS}}\leq800$ G is
due to the noise that affects the observation.}
\label{fig:fit}
\end{figure}
\begin{figure*}
\centering
\includegraphics[height=8.5cm,angle=90,clip=true]{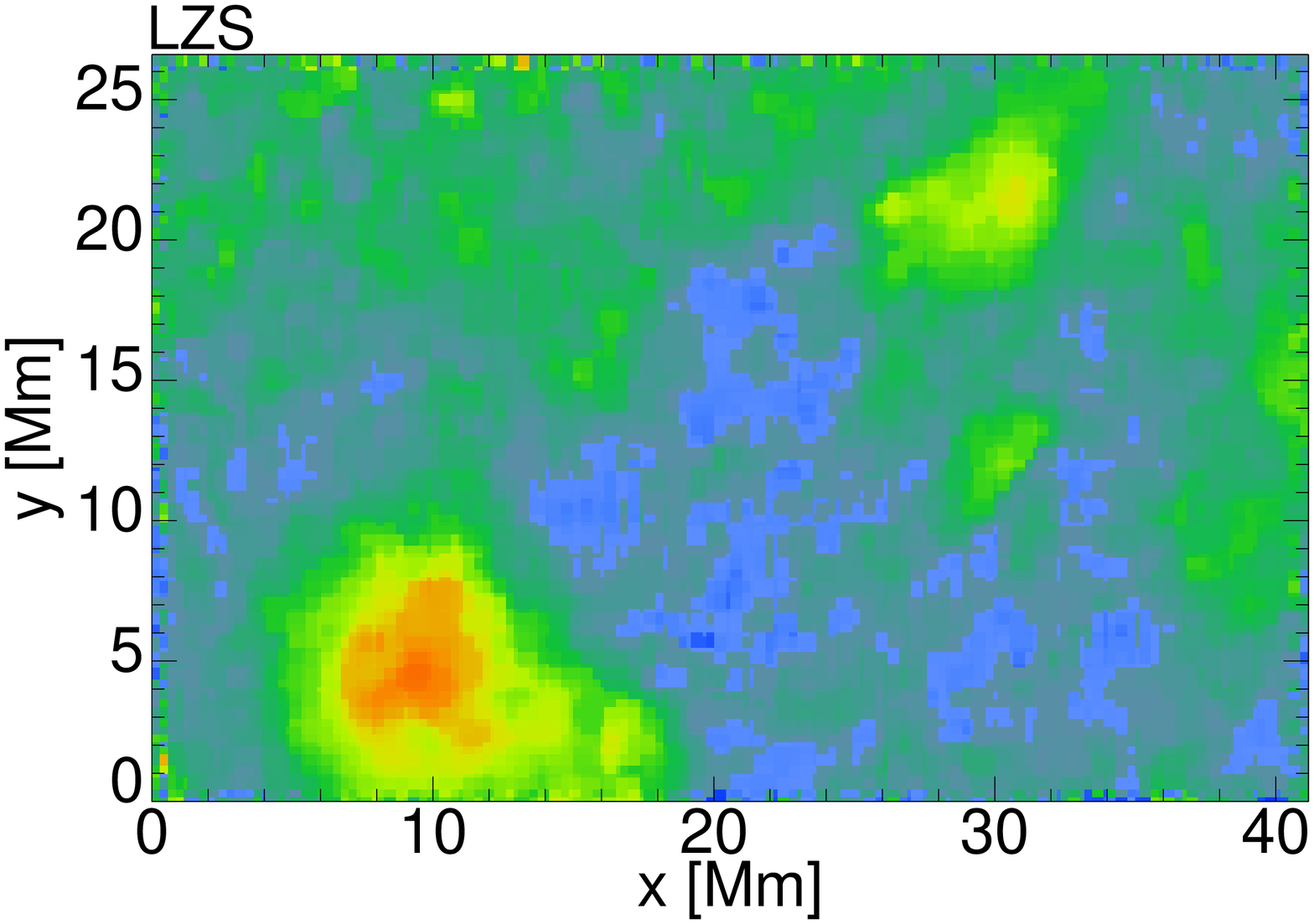}
\includegraphics[height=8.5cm,angle=90,clip=true]{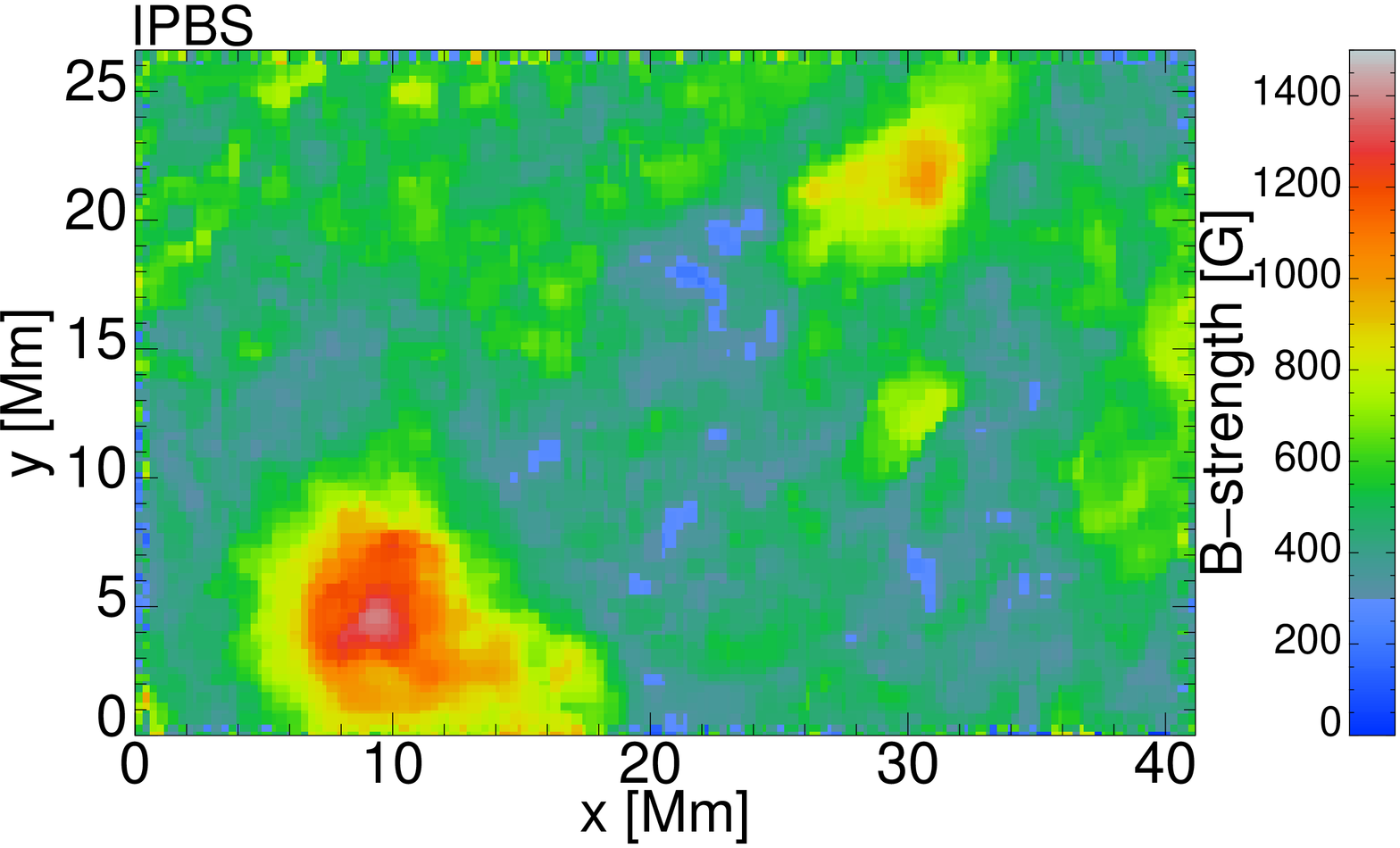}
\caption{Maps of magnetic field strength of the active region shown in
Fig.~\ref{fig:obs_int}. Left frame: result of an inversion assuming LZS;
right frame: the same, but with IPBS. Note that both images have the same
colour scale.}
\label{fig:obs}
\end{figure*}

We can now do a similar analysis using the observational data. We choose the
observations of the emerging flux region NOAA 9451 located at 33\degr W,
22\degr S that were recorded with the Tenerife Infrared Polarimeter
\citep{martinez} mounted behind the Echelle spectrograph on the Vacuum Tower
Telescope at the Teide observatory on Tenerife. Fig.~\ref{fig:obs_int}
displays Stokes $I$ maps for the observed region in the continuum and in the
core of the \ion{He}{i} line. This active region was previously studied
assuming LZS \citep{solanki_nature,lagg}. We chose this particular
observation for analysis partly because \citet{socasa} suggest that some IPBS
signatures described in their paper could be present in this observation
referring, in particular, to profile asymmetries. \\
This data set is inverted twice, once in the LZS approximation and once
including IPBS. All other particulars of the inversion remain unchanged. We
consider the error in the retrieved value of the magnetic field strength if
LZS is assumed when carrying out the inversions. Fig.~\ref{fig:fit} confirms
the result shown in Fig.~\ref{fig:pa} that the retrieved magnetic field
values are roughly 16\% higher using the IPBS inversions. This result is also
clearly displayed in the two maps of Fig~\ref{fig:obs}. It implies that the
field strengths at the chromospheric level in this region were underestimated
by on average 16\% by \citet{solanki_nature} and \citet{lagg}. We do not
expect, however, that any of the main conclusions of these publications
\citep[or of][]{Wiegelmann} are affected.

\subsection{Stokes $V$ asymmetry}

\begin{figure*}
\centering
\includegraphics[width=8.5cm,clip=true]{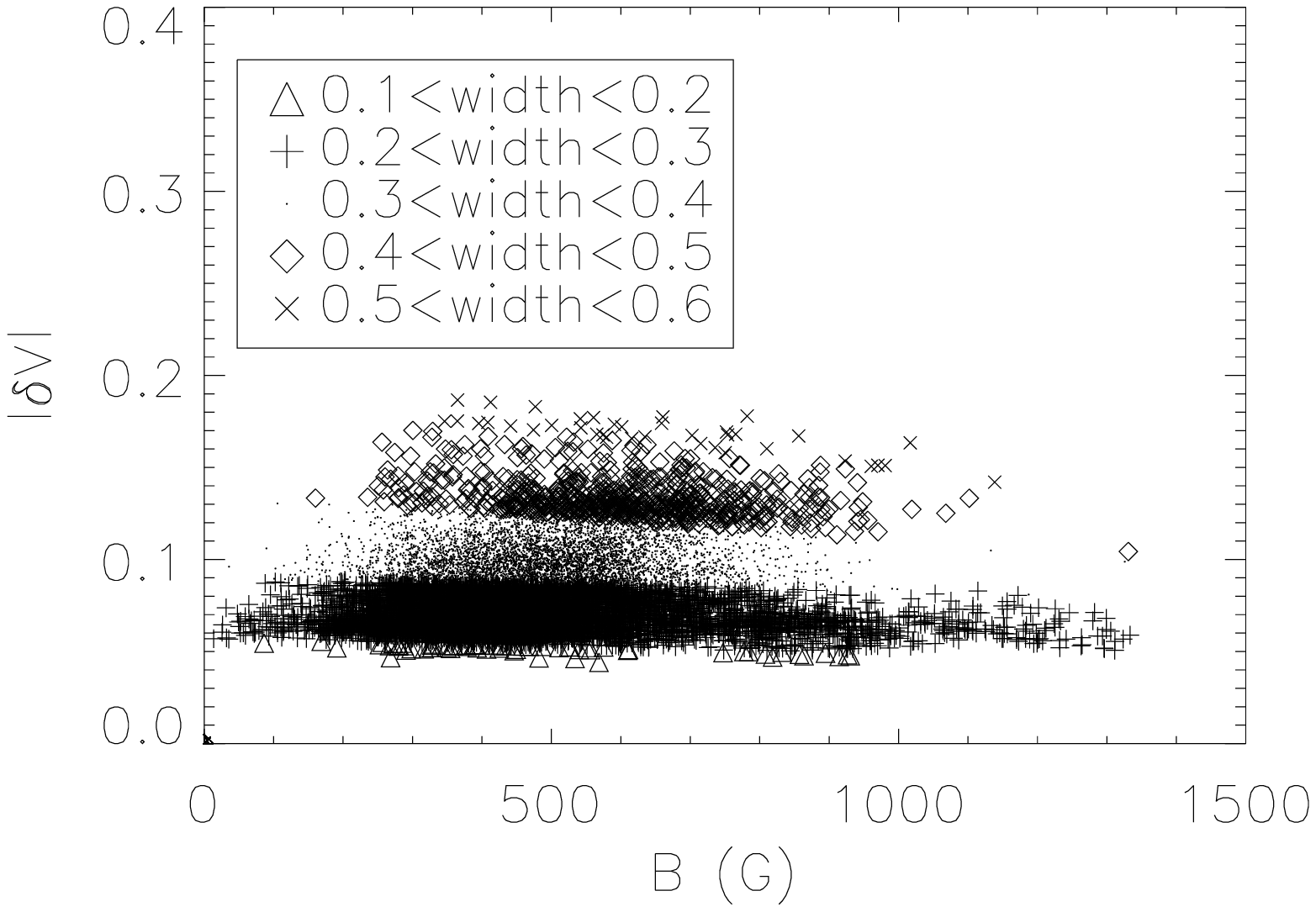}
\includegraphics[width=8.5cm,clip=true]{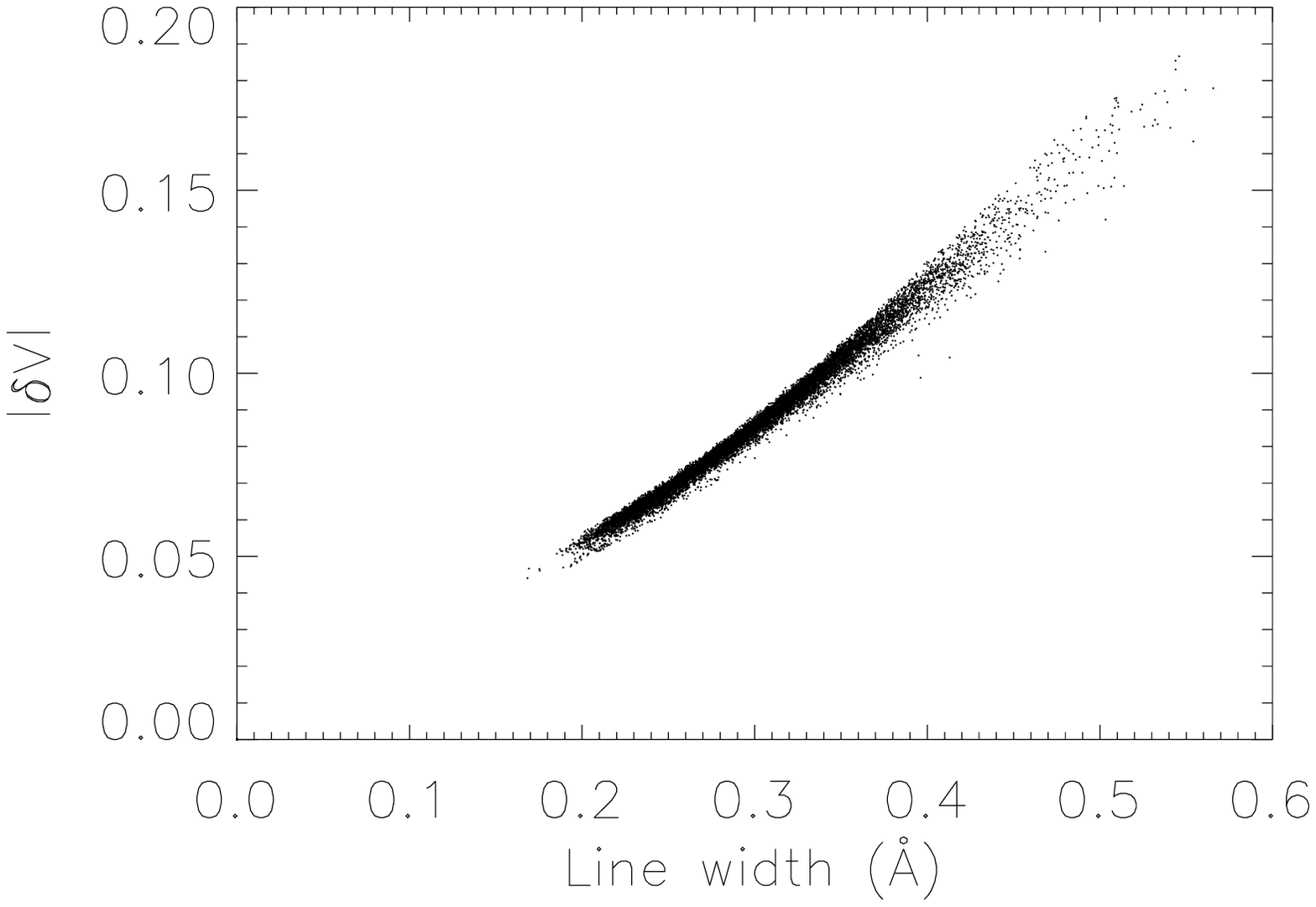}
\caption{Left: Absolute value of the relative broad-band circular
polarization ($|\delta V|$) of synthetic IPBS $V$-profiles as a function of
the magnetic field strength. The symbols distinguish between different ranges
of the e-folding width of the lines. Right: $|\delta V|$ as a function of the
width of the lines.}
\label{fig:asym1}
\end{figure*}

\begin{figure*}
\centering
\includegraphics[width=8.5cm,clip=true]{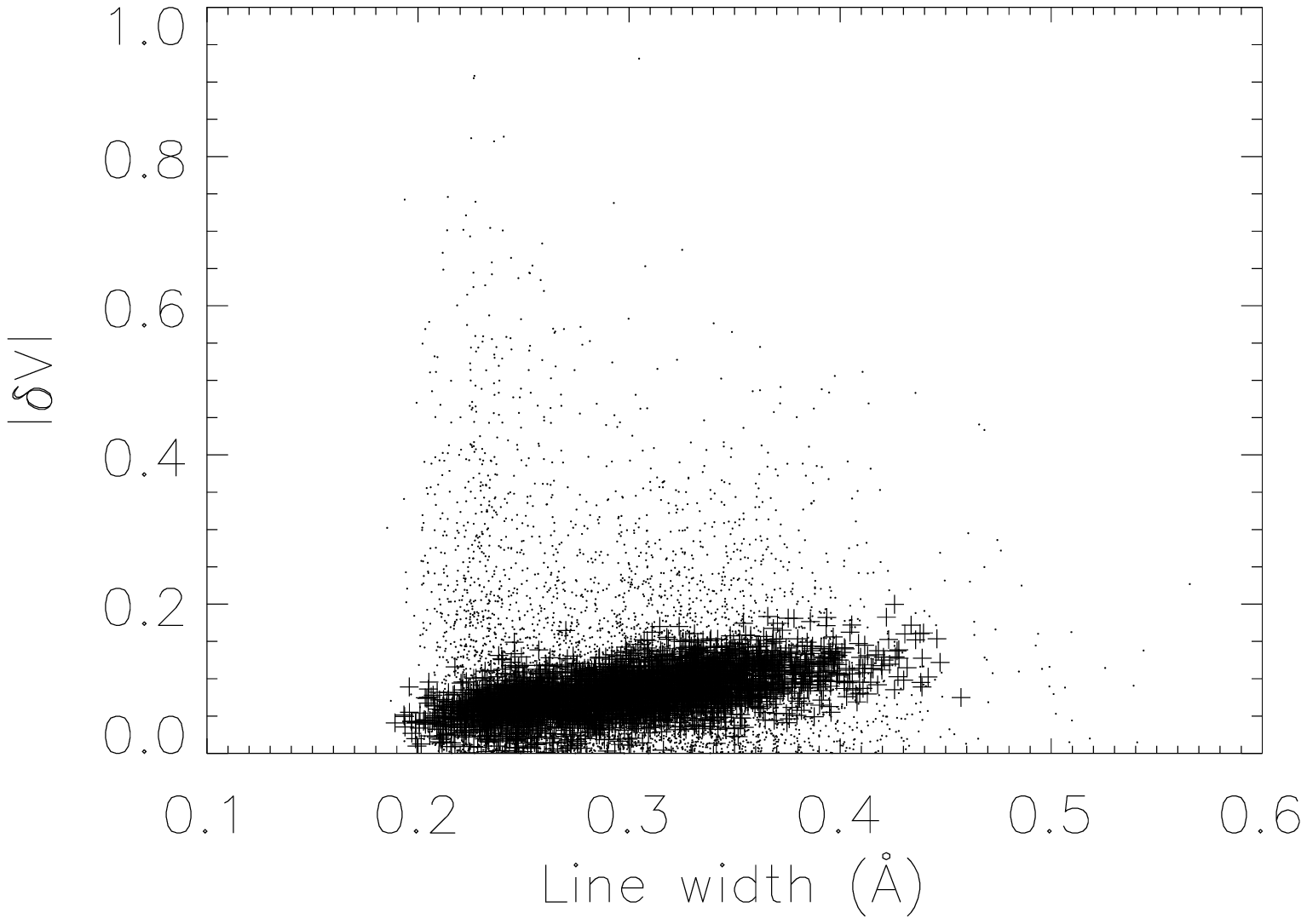}
\includegraphics[width=8.5cm,clip=true]{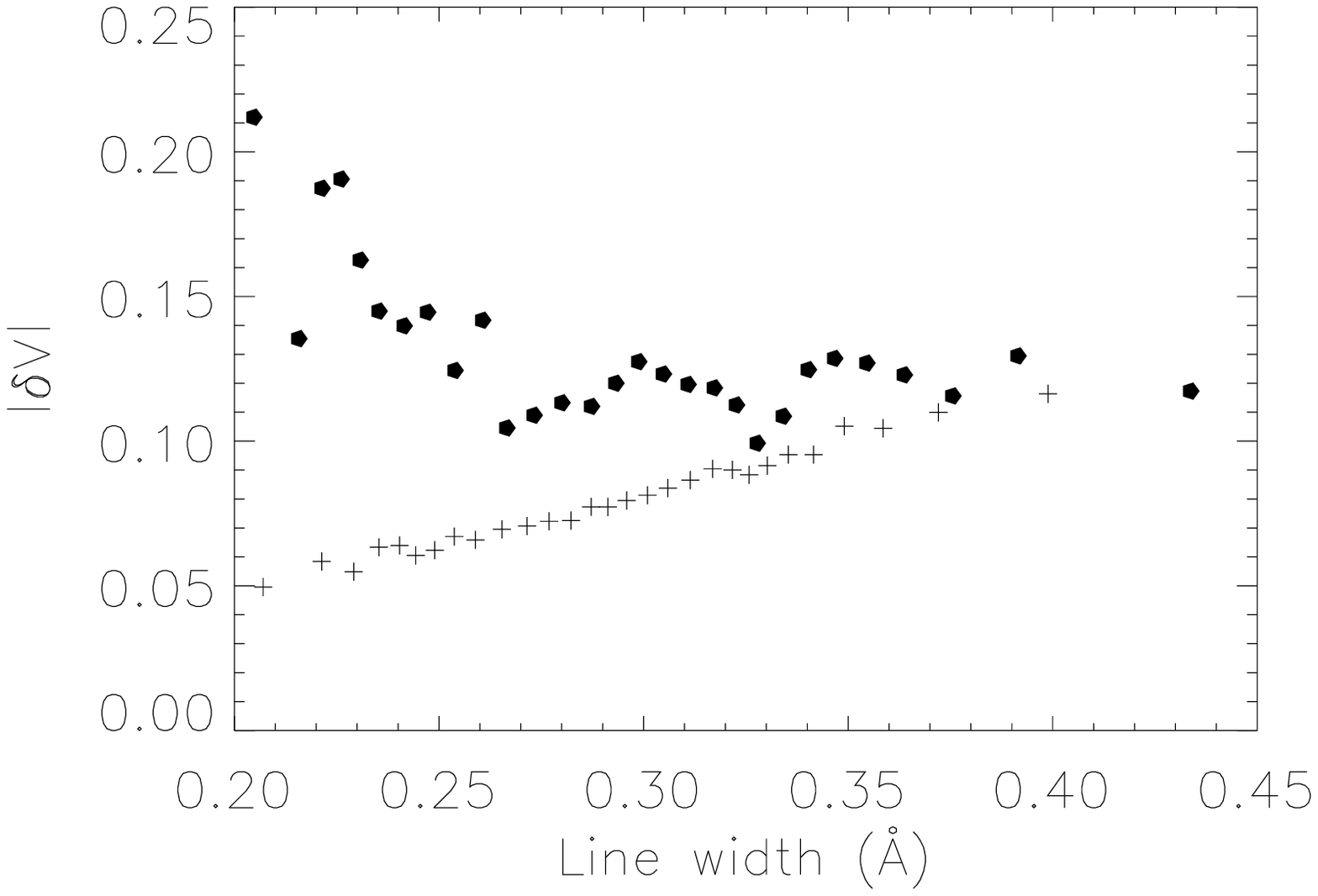}
\caption{Left: Absolute value of the area asymmetry of the fitted, synthetic
IPBS $V$-profiles (crosses) to which we have added random noise and of the
observed $V$-profiles (dots) as a function of the line width. Right: The
same, but here the values are averaged over bins containing the same number
of points.}
\label{fig:asym2}
\end{figure*}

In this section we evaluate the influence of the Paschen-Back effect on the
Stokes $V$-profile relative area asymmetry defined as \citep{solanki}
\begin{eqnarray}\label{eqn:asymmetry1}
\delta\mathcal{A}=\frac{A_b-A_r}{A_b+A_r},
\end{eqnarray}
where $A_b$ and $A_r$ (both positive) are, respectively, the areas of the
blue and the red Stokes $V$-profile wings. The parameter $\delta\mathcal{A}$
is often used to provide information on the LOS gradient of the velocity and
Bcos$\gamma$ in the height range of line formation
\citep[e.g.][]{solanki_review}. An alternative to this definition, which is
more easily applicable to a whole multiplet \citep[including complex-shaped
profiles;][]{lagg1}, is the relative broad-band circular polarization
\begin{eqnarray}\label{eqn:asymmetry2}
\delta V=\frac{\int_{-\infty}^{+\infty}V(\lambda)d\lambda}{\int_{-\infty}^{+\infty}|V(\lambda)|d\lambda}.
\end{eqnarray}
\citet{socasa} investigated how the area asymmetry
produced by the Paschen-Back effect varies with the magnetic field strength
and pointed out that for a typical Doppler width of 105 m{\AA},
$|\delta V|\approx0.11$ for $B<1000$ G, which suggests that significantly
larger observed values of $|\delta V|$ cannot be solely due to the effects of
Paschen-Back effect splitting if the inferred B turns out to be smaller or
similar to 1000 G. \\
In Fig.~\ref{fig:asym1} (left panel) we plot the absolute value of the
relative broad-band circular polarization ($|\delta V|$) of the best-fit IPBS
$V$-profiles to the observed profiles as a function of the magnetic field
strength for different ranges of the values of the e-folding width of the
lines. The $\delta V$ is calculated over the whole wavelength range of the He
multiplet. The $|\delta V|$ of the best-fit synthetic IPBS $V$-profiles is
nearly independent of B, but is a strong function of the width of the lines,
as indicated by different symbols. This behaviour is displayed more clearly
in the right panel of Fig.~\ref{fig:asym1} where we plot $|\delta V|$ as a
function of the width of the lines. We expect that the $|\delta V|$ of the
observed profiles should exhibit a similar dependence, if $|\delta V|$ solely
results from the IPBS. The $|\delta V|$ of the observed profiles (see dots in
Fig.~\ref{fig:asym2}, left panel) clearly shows no such dependence. In order
to test if noise is the reason for the lack of this dependence we add random
noise to the synthetic IPBS $V$-profile of the same order as the one we have
in the observational data. We restrict our analysis to synthetic profiles
with a maximum $V$-signal at least three times higher than the noise level
and to the corresponding observed profiles. The $|\delta V|$ of synthetic,
noisy IPBS $V$-profiles (represented by crosses in the left panel of
Fig.~\ref{fig:asym2}) retains the dependence on the width of the lines.
Fig.~\ref{fig:asym2} (left) shows that the observed $|\delta V|$ values can
be up to 4 times as large as those of the synthetic profiles. Values of the
observed $|\delta V|\geq0.2$ cannot be explained by IPBS or noise. The
difference between the two sets of $|\delta V|$ values is displayed more
markedly in the right panel of Fig.~\ref{fig:asym2}, where the values are
grouped into bins containing the same number of points. For each bin the mean
value is displayed in the x- and y-coordinates. Moreover, the difference
between the $\delta V$ values of the synthetic IPBS $V$-profiles and of the
observed $V$-profiles is even larger than suggested by Fig.~\ref{fig:asym2}
because, for 54\% of the profiles, they have opposite sign. Consequently,
another mechanism must be acting to produce at least the larger values of the
asymmetry. The most obvious one is the combination of velocity and
magnetic-field gradients \citep[see][for a review]{solanki_review}. Support
for this mechanism comes from the fact that the observed $\delta V$ is larger
for narrow lines, for which smaller velocity and magnetic-field gradients are
required to produce a large $\delta V$ \citep{grossmann}.

\section{Conclusions}

We analysed the influence of the Paschen-Back effect on the Stokes profiles
of the \ion{He}{i} 10830 {\AA} multiplet lines, estimating its relevance
using synthetic profiles and investigating its influence on the inversion of a
spectropolarimetric scan of an emerging active region. \\
Our results support the conclusion of \citet{socasa} that IPBS should
be taken into account when modeling the polarization profiles of the
\ion{He}{i} 10830 multiplet. For example, we found that by including the
incomplete Paschen-Back effect into our inversion code, on average 16\%
higher field strength values are retrieved from inversions, while other
atmospheric parameters are affected less significantly. \\
We also showed that the Paschen-Back effect is not the main cause for the
area asymmetry exhibited by the many \ion{He}{i} 10830 Stokes profiles
that showed a line width smaller than $\sim0.30$ {\AA} in Lagg et
al.'s (2004) emerging flux region observation. The spatial points
corresponding to such Stokes $V$-profiles are those for which the inferred
value of the magnetic field strength is lower than $\sim1.1$ kG
when IPBS is taken into account. The fact that the area asymmetry of the
observed $V$-profiles is considerably stronger than of the synthetic
$V$-profiles indicates that some other effect drives the area asymmetry more
strongly than the Paschen-Back effect. The main candidates are LOS-gradients
of the magnetic field vector and the velocity.

\begin{acknowledgements}
We would like to thank Hector Socas-Navarro and Javier Trujillo Bueno for
helpful discussions regarding the Paschen-Back effect and for providing
material from their publications. We also thank Joachim Woch, Norbert Krupp
and Manolo Collados for their help during the observing run. This work was
supported by the program ``Acciones Integrades Hispano-Alemanas'' of the
German Academic Exchange Service (DAAD project number D/04/39952).
\end{acknowledgements}

\bibliographystyle{aa}

\end{document}